\documentstyle[multicol,floats,epsfig,aps]{revtex}

\begin{document}
\draft

\title{Magnetostructural effects and phase transition \\
in Cr$_2$O$_3$ under pressure}
\author{Alexander Yu. Dobin, Wenhui Duan, and Renata M. Wentzcovitch}
\address{Department of Chemical Engineering and Material Sciences,
         School of Physics and Astronomy,\\
	 and Minnesota Supercomputer Institute,
	 University of Minnesota, Minneapolis, MN 55455}
\date{\today}

\maketitle

\begin{abstract}

We have successfully calculated the electronic and structural properties
of chromia (Cr$_2$O$_3$) in the Local Spin Density Approximation (LSDA).
We predict a transformation from the corundum to the Rh$_2$O$_3$~(II)
structure around 15 GPa in the anti-ferromagnetic
(AFM) phase as well as in the paramagnetic (PM) insulating 
state which occurs above the N\'eel temperature ($T_N$).
This transition is relevant to interpreting 
the optical anomalies observed in the absorption spectrum of 
ruby under pressure.
We have modeled the structural properties of the
PM state using a Landau-like expansion of the magnetostriction
energy. This treatment correctly describes the structural anomalies 
across $T_N$ in the corundum phase and indicates that the AFM and
PM insulating states should have distinct compressive behaviors.

\end{abstract}

\begin{multicols}{2}
%
%
Oxides of the $3d$ transition metals are a fascinating class of materials
with amazingly diverse physical properties. They have been a subject
of intensive experimental and theoretical studies for many years
\cite{mott49,terakura84,vo2,lewis97}.
A wide variety of computational techniques --- density-functional theory
with linear augmented plane wave method
\cite{terakura84,dufek94}, non-periodic embedded cluster approach
\cite{casanovas94,mejias95}, periodic unrestricted Hartree-Fock
method \cite{towler94,catti96} --- have been used to perform 
{\it ab initio} calculations of these compounds. 
In this paper we report a successful application of the first principles 
pseudo-potential plane-wave (PPPW) approach \cite{cohenpick} to
compute structural properties of and predict a pressure induced 
structural transition in chromia ($Cr_2O_3$), a typical 
anti-ferromagnetic insulator in this class of materials.
There were a few experimental papers on chromia 
\cite{lewis66,sato79,finger80,henrich92} and only one 
{\it ab initio} calculation \cite{catti96}. The periodic unrestricted 
LCAO Hartree -- Fock method was used \cite{catti96}
to calculate crystal parameters and elastic properties, as well as 
electronic structure of $Cr_2O_3$. Their results are in a good agreement 
with experimental data.
The advantage of the PPPW approach is in the simplicity of
the plane-wave basis set, which makes it easy to calculate 
ionic forces and lattice stresses. This allowed us to optimize 
dynamically the cell structure of chromia under pressure\cite{fpvcs} for 
various magnetic states. 

Our motivation to study $Cr_2O_3$ is related to the high pressure
behavior of ruby, i.e. $Al_{2(1-x)}Cr_{2x}O_3 \quad (x < 0.05)$.
The pressure dependence of the fluorescence lines in ruby is 
widely used to determine pressure (the so-called ruby scale) 
in diamond-anvil-cell experiments.
Alumina ($Al_2O_3$) and chromia exist in the corundum phase and
form a completely isomorphous alloy system.
Alumina has been shown to undergo a structural phase transition
to the $Rh_2O_3$~(II) phase around 80-100 GPa \cite{alo1,cohen,funajean}.
A recent theoretical study\cite{ruby1} of the effect of this structural
transition on the optical spectrum of ruby indicated that the 
neighborhood of the chromium site, a {\it color center}, might be undergoing 
a severe distortion around 30 GPa.
This hypothesis is suggested by the behavior of the optical absorption 
lines which display a small discontinuity at 30 GPa and 
resemble more closely the transitions predicted in the
high pressure $Rh_2O_3$~(II) phase beyond 30 GPa than those in the corundum
phase. It was then anticipated that the cause of this distortion
could be a similar phase transformation
at lower pressures ($<$ 30 GPa) in chromia, the other
end member of the alloy. At the moment there is no convincing 
experimental evidence for a phase transformation
in $Cr_2O_3$ in this pressure range \cite{liubass}. 
However, this may be due to the fact that the $Rh_2O_3$ (II) phase has an 
X-ray diffraction pattern similar to corundum's \cite{alo1}, 
which makes it difficult to observe.

In this paper we investigate this pressure 
induced transformation in chromia. However another interesting 
question arises: at room temperature chromia undergoes a change 
in magnetic phase under pressure. Its N\'eel temperature is
$T_N=308$ K \cite{corliss65}, with a pressure dependence of 
$\partial T_N/\partial P=-16 \ K/GPa$ \cite{worlton68}.
The paramagnetic (PM) state above $T_N$ is also insulating; 
therefore, the effect of the magnetic transition on the structural 
properties are not expected to be dramatic. Nevertheless, 
structural anomalies around the N\'eel temperature are well known 
\cite{alberts76,greenwald56} and a realistic prediction of a possible
phase transition above 0.5 GPa should be carried out in the PM 
insulating phase with randomly oriented spins. 
Here we investigate from first principles this structural 
transition in the AFM, and in a hypothetical 
ferromagnetic (FM) phases. The structural properties of
PM insulating state is then explored in relation to those of
the AFM and FM phases using a phenomenological 
approach based on a Landau-like expansion of the magnetostriction 
energy. The predicted structural differences of this phase
with respect to the AFM phase correlate well with the anomalies 
\end{multicols}
\twocolumn
\begin{figure}
\epsfig{file=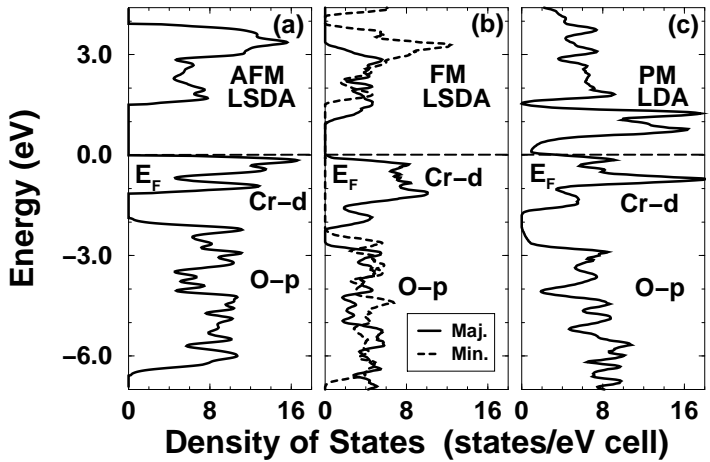,width=3.4in}
\caption{Calculated DOS for various magnetic phases 
of Cr$_2$O$_3$ in the corundum structure.}
\label{fig_bands}
\end{figure}
\noindent
observed around the N\'eel transition and are very different 
from those of the PM metallic phase predicted by a standard
LDA calculation.

The crystal structure of $Cr_2O_3$ at ambient conditions is corundum-like.
It can be described as a hexagonal closed packed array of oxygens with
two thirds of the octahedral sites filled with chromium atoms. The unit cell
is rhombohedral and contains two formula units.
In  $Cr_2O_3$ each chromium is left with three $d$-electrons, losing 
the other three to oxygens. The predominantly octahedral crystal field 
splits $d$-orbitals into (approximately) a $t_{2g}$-like  
triplet and an $e_g$-like doublet. The lower triplet accommodates
three electrons. The $Cr^{3+}$ local magnetic moment is $2.76 \mu_B$, \cite{corliss65}
close to the spin only value of $3 \mu_B$.
The moments lie parallel to the $c$-axis, and are arranged
in ferromagnetic layers perpendicular to the $c$-axis.
The layers alternate up and down along the $c$-axis. In the 
insulating phase above $T_N$ chromia has randomly oriented spins.

To investigate the effect of magnetism on the structural properties 
we have performed three distinct calculations:
1) a standard {\it spin-polarized} LSDA calculation \cite{descr_method} 
in the AFM phase of $Cr_2O_3$ in the corundum structure;
2) same-type calculation in the FM phase, with the net magnetic moment 
of $3 \mu_B$ per $Cr$~atom;
3) a standard {\it non-spin-polarized} LDA calculation in a PM phase. 
In all cases the lattice and internal
degrees of freedom were dynamically relaxed under 
pressure \cite{fpvcs}. The zero-pressure structures obtained 
correspond to various local minima of the LSD functional.

The ground state at T = 0 K is the AFM state with a band
gap of $\approx$ 1.5 eV (Fig.~\ref{fig_bands}-a)
and a local magnetic moment of $3\mu_B$ on chromium atoms
(from straight band
occupations). 
The overall band structure compares well with 
photo-emission data \cite{wertheim73,eastman75}. Namely,
the O$_{2p}$ and Cr$_{3d}$ band widths of 5 eV and 1 eV
respectively, and O$_{2p}$-Cr$_{3d}$ band centers separation
of 4 eV are in a good agreement with experimental values.
However, as expected, the band gap is underestimated with respect
to the thermal gap of 3.3 eV \cite{c64}.  
Zero-pressure equilibrium structural parameters presented 
in Table \ref{table1} are also in good agreement with experimental 
data. The cohesive energy of E$_{coh}$ = 6.1 eV/atom compares well 
with $E_{coh} = 5.55$ eV/atom from experiments. 
The FM state is found to be insulating as well with a band gap 
of $\approx$~0.9~eV (Fig.\ref{fig_bands}-b). After structural
relaxation this state is only $35$~meV/unit above the AFM ground state.
Equilibrium lattice parameters (Table~\ref{table1}) are
quite different from those in the AFM state, indicating a substantial
influence of the magnetic state on the structural properties.
The standard 
paramagnetic {\it non-spin-polarized} LDA calculation stabilizes
$Cr_2O_3$ in a {\it metallic} phase~(Fig.\ref{fig_bands}-c),
as expected, with structural properties considerably different from 
the observed ones (see Table \ref{table1}). This state is 
2.25 eV/unit above the AFM ground state and cannot properly account 
for the structural transition under consideration. 

\begin{table}[!t]
\caption{Zero pressure structural parameters of chromia in the corundum 
structure (rhombohedral unit cell): lattice constant $a_0$ (\AA) and rhombohedral
angle $\alpha$ (deg), internal atomic coordinates $u(Cr)$ and $u(O)$,
bulk modulus $B_0$ (GPa) and its pressure derivative $B_0'$.} \label{table1}
\begin{tabular}{ccccccc}
                  & AFM LSDA  & FM LSDA  & PM LDA    & Experiment\\ 
\hline
$a_0$        &$5.366$    &$5.308$  &$5.688$    & $5.362$  \tablenotemark[3] \ $5.350$  \tablenotemark[4] \\ 
$\alpha$   &$55.17$    &$56.14$  &$47.23$    & $55.108$ \tablenotemark[3] \ $55.128$ \tablenotemark[4] \\  

$u(Cr)$     &$0.347$    &$0.351$  &$0.337$    & $0.3475$ \tablenotemark[3] \ $0.3477$ \tablenotemark[4] \\
$u(O)$      &$0.557$    &$0.550$  &$0.583$    & $0.556$  \tablenotemark[3] \ $0.555$  \tablenotemark[4] \\
$B_0$\ \tablenotemark[1]
           &$251\pm6$   & $215\pm5$ &$300\pm8$  & $238\pm4$\tablenotemark[4] \ $222\pm2$ \tablenotemark[6]  \\
$B_0$\ \tablenotemark[2]
           &$261\pm2$  &$211\pm5$   &$297\pm3$  & $231\pm5$\tablenotemark[6]   \\
$B_0'$\   \tablenotemark[2]    
           &$2.59$         &$2.73$           &$4.24$     &    $2.0\pm1.1$\tablenotemark[6]  \\
\end{tabular}
\tablenotetext[1]{Second order finite strain equation of state (FSEoS)
($B_0'\equiv $4).
We used our data up to 15 GPa for this fitting.}
\tablenotetext[2]{Third order FSEoS was used with $B_0'$ as free parameter.
We include pressures up to 140 GPa to get a correct value of $B_0'$, while
in Ref.~\cite{sato79} the pressure range was not sufficient for a confident
determination of $B_0'$.}
\tablenotetext[3]{Ref. \cite{lewis66}}
\tablenotetext[4]{Ref. \cite{finger80}}
\tablenotetext[6]{Ref. \cite{sato79}}
\end{table}

Our description of the structural properties of the PM phase
is based on the Landau-like expansion of the crystal deformation energy
at a certain fixed pressure,
$
E =  \lambda_{ikjl} u_{ik} u_{jl} + \beta_{ikjl}u_{ik}M_{1j}M_{2l}.
$
Here $u_{ik}$ is a strain tensor, 
${\bf M}_1$, ${\bf M}_2$ are the magnetization vectors of the 
two AFM sublattices, and 
$\bf \lambda$ and $\bf \beta$ are constant tensors.
The first term represents pure elastic deformation,
while the second describes the magnetostriction energy, i. e., the
coupling of the magnetic and structural degrees of freedom. 
Deformations conserving the corundum structure symmetry
allow only for $U=u_{33}$, uniaxial strain, and $V=\sum{u_{ii}}$, 
hydrostatic compression. We assume ${\bf M}_1$, ${\bf M}_2$
remain parallel to the $z$-axis.
The above expression then simplifies to:
$
E=  \lambda_1 V^2 + \lambda_2 U^2  -  \lambda_{12}VU
                             +(\beta_1 V + \beta_2 U) \cdot M_{1z} M_{2z}.
$ 
The equilibrium configurations are such that
$U=u_0 \cdot M_{1z} M_{2z}$, $V=v_0 \cdot M_{1z} M_{2z}$, 
where $u_0$ and $v_0$ depend on $\lambda$ and $\beta$.
For the AFM and FM phases $M_{1z} M_{2z}=-1$ and  $M_{1z} M_{2z}=1$ 
respectively. For the PM state the average product 
$\langle M_{1z} M_{2z} \rangle_{PM}$ = 0.
Therefore, the equilibrium lattice structural parameters of the PM 
phase are given by:
$U_{PM}=(U_{AFM}+U_{FM})/2$, $V_{PM}=(V_{AFM}+V_{FM})/2$
where $U_{AFM}$, $U_{FM}$, $V_{AFM}$, and $V_{FM}$ have been 
determined from first principles. A similar procedure can be
adopted for dealing with the Rh$_2$O$_3$ (II) phase. In this 
case, the deformation energy is expressed as:
$E =  \sum \lambda_{\gamma\delta} u_{\gamma}u_{\delta}
 +\sum \beta_{\gamma} u_{\gamma} \cdot M_{1z} M_{2z}$,
where $\gamma, \delta =xx, yy, zz$.

The predicted properties of the AFM and PM insulating states 
compare as follows: a) the difference in zero pressure lattice parameters 
(in the hexagonal cell description) between them, are similar 
to the anomalies observed around $T_N$. Throughout the N\'eel
transition (AFM to PM), the calculated $\Delta a_H = +0.013\AA$ and 
$\Delta c_H = -0.11 \AA$ agree in sign and approximately in order of magnitude
with experimental values of 
$\Delta a_H = +0.006 \AA$ and $\Delta c_H =
-0.018 \AA$ \cite{alberts76}. We believe this is evidence of the satisfactory
description of the PM insulating state.
b) The calculated compressive behavior of our PM insulating state 
compares well with the experimental behavior under pressure above
T$_N$ \cite{sato79}, while our AFM calculation is in better agreement
with the experiment of Lewis and Drickamer\cite{lewis66}, 
which we suspect may have been carried out in the AFM phase.

\begin{figure}[!t]
\vskip 3mm
\epsfig{file=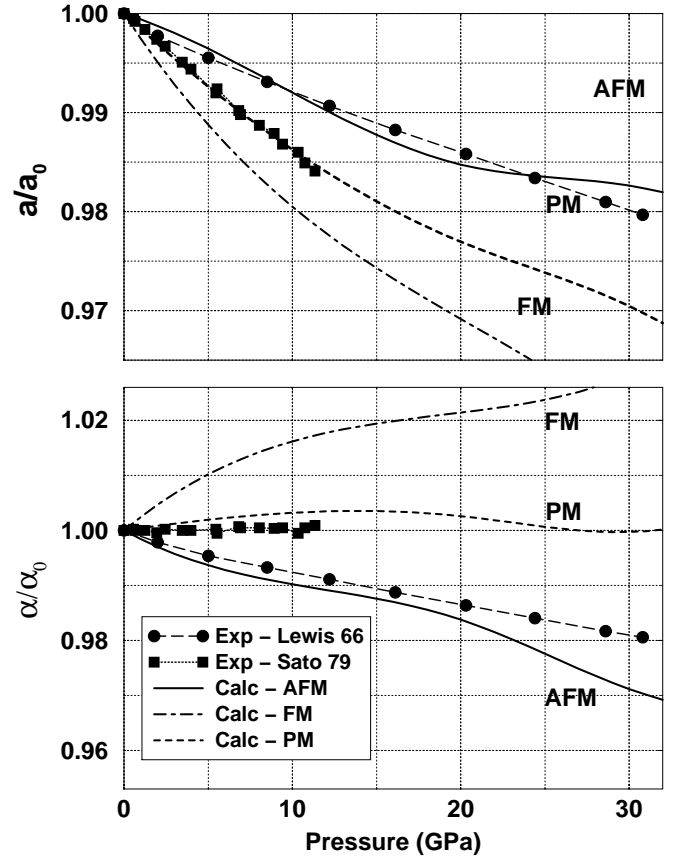,width=3.4in}
\caption{Pressure dependence of the $Cr_2O_3$ rhombohedral cell constant $a$ and
angle $\alpha$, calculated for AFM, FM and PM phases compared to experimental
measurements from Ref.~\protect\onlinecite{lewis66} (Lewis 66) 
and Ref.~\protect\onlinecite{sato79} (Sato79).}
\label{fig1}
\end{figure}

A summary of experimental data and our results is presented in 
Fig.~\ref{fig1}. In Ref.~\onlinecite{lewis66} a substantial decrease 
of the rhombohedral angle with pressure was found (hexagonal $c$-axis 
less compressive than $a-b$ axes). In contrast,
in Ref.~\onlinecite{sato79} a slight increase in this
angle was observed ($c$-axis more compressive than $a-b$ axes). 
This discrepancy has been attributed to non-hydrostatic stresses 
in Ref. \onlinecite{lewis66}. Our results suggest this discrepancy 
could be real if somehow in Ref. \onlinecite{lewis66} chromia was kept 
in the AFM state. We predict here that the magnetic state affects 
noticeably the compressive behavior of chromia despite the 
insulating nature of both phases. Compression experiments well 
below and well above $T_N$ could help to clarify this situation.

\begin{figure}[!t]
\vskip 5mm
\epsfig{file=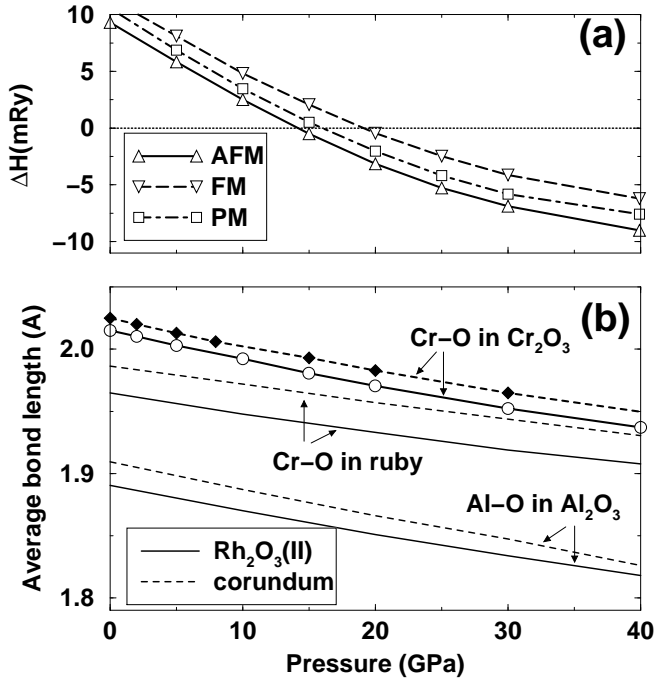,width=3.4in}
\caption{{\bf (a)} \ Pressure dependence of the enthalpy for 
Rh$_2$O$_3$~(II) structure, relative
to the corundum in AFM and PM phases;\ \ 
{\bf (b)} \ Average radii of the first
coordination shell around chromium in chromia and ruby \protect\cite{ruby1},
and aluminum in alumina \protect\cite{ruby1}.}\label{fig4}
\end{figure}

Now we deal with the structural transition. 
The high pressure $Rh_2O_3$~(II) 
phase has $Pbna$ space group with an orthorhombic unit cell
containing 20 atoms (four Cr$_2$O$_3$ units). This phase is structurally
similar to the corundum structure and may be described as having
a different stacking of similar octahedral layers with a periodicity
which is twice that of the corundum along the '$c$'
direction\cite{shannon70}.
In the corundum structure the CrO$_6$ octahedra share three edges 
while in the Rh$_2$O$_3$~(II) structure they share only two.
The relative stability of these structures in both magnetic states is
shown in Fig. \ref{fig4}-a. We predict the corundum to $Rh_2O_3$~(II) 
transformation in chromia to take place at 14 GPa and 16 GPa 
in the AFM and in the PM insulating phases respectively, with fractional 
volume changes at the transition of $\approx -2\%$.
This transition would also take place in the FM phase if it were 
somehow stabilized.

Fig. \ref{fig4}-b displays the average radius of the first
coordination shell around chromium in AFM chromia and in ruby
(from Ref.~\onlinecite{ruby1}). The average Cr-O bond-length 
in chromia increases across the corundum to Rh$_2$O$_3$~(II) transition
by an amount similar to that required to explain the optical 
anomalies in ruby under pressure, which can be explained by
a decrease in crystal field splitting. This verifies that this
presumable rearrangement could arise from a preference of chromia 
for the Rh$_2$O$_3$~(II) phase above $\approx$ 15 GPa. The structural
constraint imposed by the alumina host structure should 
naturally hinder the atomic rearrangement around the color 
centers until higher pressures, for instance 30 GPa. 

These results should stimulate further experimental and theoretical work.
The prediction of distinct compressive behaviors in chromia above
and below T$_N$ and the structural phase transformation near 15 GPa
await experimental confirmations.
The latter, if verified, makes ruby an interesting study
case: an isomorphous alloy in which both end members undergo the
same structural transition but at very different pressures.
Intermediate compositions should undergo similar transitions at
intermediate pressures. However, before the transformation manifests
macroscopically it could be nucleating around one of the components, 
even in the impurity limit. The possibility of investigating this
phenomenon in ruby by EXAFS or anomalous X-ray scattering is fascinating.

\noindent{{\it Acknowledgments} - We thank Phil Allen and Stefano Baroni 
for helpful
comments on the manuscript. This work was supported by the University of
Minnesota MRSEC (AYuD), and NSF grant EAR-9973130 (RMW and WD).}

\begin{references}
\bibitem{descr_method} 
The norm-conserving Troullier-Martins
\cite{TM} pseudopotentials (cutoff radii in a.u.: $O$ $r_s=r_p=1.45$;
Cr $r_s=r_p=2.5$, $r_d=1.75$) with partial core correction \cite{corecorr}
(pseudocore radius for $Cr$: $r_c=1.6$) in the plane-wave basis set
were used \cite{cohenpick}.
The plane-wave energy cutoffs were 70 Ry for
wave functions and 280 Ry for charge densities and potentials.
Brillouin-zone summations were performed over two k-points in the
irreducible wedge. We used Ceperley and Alder LSDA 
exchange-correlation potential
parameterized by Perdew and Zunger \cite{CA80}.
\end{references}
\end{document}